\documentclass[aps,prc,12pt,showpacs,showkeys,onecolumn,superscriptaddress,groupedaddress]{revtex4-1}
\usepackage{graphicx}
\usepackage{graphics}
\usepackage{xspace}
\usepackage{amssymb}
\usepackage{amsmath}
\usepackage[dvips]{color}
\usepackage{latexsym}
\usepackage{mathtools}
\usepackage{mathrsfs}
\usepackage{url}
\usepackage[english]{babel}
\usepackage{color}
\usepackage{bm}
\newcommand{\inieq}{\begin{eqnarray}}            
\newcommand{\fineq}{\end{eqnarray}}            
\newcommand{\diff}{{\rm\,d}}                    
\newcommand{\bint}{\mskip .5mu \int \mskip-18mu} 
\newcommand{\be}{\begin{equation}}
\newcommand{\ee}{\end{equation}}
\newcommand{\ba}{\begin{eqnarray}}
\newcommand{\ea}{\end{eqnarray}}

\def\r{\mbox{{\bf  r}}}

\def\q{\mbox{\boldmath $q$}}

\def\ee{\mbox{$\left(e,e^{\prime}\right)$\ }}

\def\mcv{\mbox{$\mathcal{V}$}}
\def\caq{\mbox{$^{48}$Ca}}
\def\pbd{\mbox{$^{208}$Pb}}
\DeclarePairedDelimiter{\abs}{\lvert}{\rvert}
\DeclareMathOperator{\Realpart}{Re}
\renewcommand{\Re}{\Realpart}
\DeclareMathOperator{\Imagpart}{Im}
\renewcommand{\Im}{\Imagpart}

\begin{document}
\title{Elastic and quasi-elastic electron scattering off nuclei with neutron excess}
\author{Andrea Meucci} 
\affiliation{Dipartimento di Fisica, 
Universit\`{a} degli Studi di Pavia and \\
INFN,
Sezione di Pavia, via A. Bassi 6, I-27100 Pavia, Italy}
\author{Matteo Vorabbi}
\affiliation{Dipartimento di Fisica, 
Universit\`{a} degli Studi di Pavia and \\
INFN,
Sezione di Pavia, via A. Bassi 6, I-27100 Pavia, Italy}
\author{Paolo Finelli}
\affiliation{Dipartimento di Fisica e Astronomia, 
Universit\`{a} degli Studi di Bologna and \\
INFN,
Sezione di Bologna, via Irnerio 46, I-40126 Bologna, Italy}
\author{Carlotta Giusti}
\affiliation{Dipartimento di Fisica, 
Universit\`{a} degli Studi di Pavia and \\
INFN,
Sezione di Pavia, via A. Bassi 6, I-27100 Pavia, Italy}
\author{Franco Davide Pacati}
\affiliation{Dipartimento di Fisica, 
Universit\`{a} degli Studi di Pavia and \\
INFN,
Sezione di Pavia, via A. Bassi 6, I-27100 Pavia, Italy}
\date{\today}

\begin{abstract}
We present theoretical predictions for  
electron scattering  on oxygen  and calcium  isotopic 
chains. 
The calculations are done within the framework of the distorted-wave
Born approximation and the proton  and neutron density distributions are 
evaluated adopting a relativistic Dirac-Hartree model. We present results 
for the elastic and quasi-elastic cross sections and for the parity-violating asymmetry.
As a first step, the results of the models are tested in comparison with some of 
the data available for elastic and quasi-elastic scattering on $^{16}$O and 
$^{40}$Ca nuclei. Then, the evolution of some nuclear properties is investigated 
as a function of the neutron  number. 
We also present a comparison with the parity-violating asymmetry parameter 
obtained by the PREX Collaboration on \pbd\ and give a prediction for the 
future experiment CREX on \caq.

 \end{abstract}


\maketitle


\section{Introduction}
\label{intro}

The nuclear response to an external probe is a powerful tool to investigate 
the structure of hadron systems such as atomic nuclei and their constituents.
In particular, electron-scattering reactions 
have provided the most complete and detailed information on nuclear and nucleon
structure
\cite{Hofstadter:1956qs,Donnelly:1975ze,Donnelly:1984rg,Boffi:1993gs,book,RevModPhys.80.189}. 
Electrons interact with nuclei predominantly through the electromagnetic 
interaction, which is well known from quantum electrodynamic  and is weak 
compared with the strength of the interaction between hadrons. 
The scattering process is therefore adequately treated assuming the validity of 
the Born approximation, i.e., the one-photon exchange mechanism between 
electron 
and target. The virtual photon, like the real one, has a mean free path much
larger than the target dimensions, thus exploring the whole target volume. 
This contrasts with hadron probes, which are generally absorbed at the target 
surface. 
Moreover, the ability to vary independently the energy $\omega$ and momentum
{\boldmath${q}$}
of the exchanged virtual photon transferred to the nucleus makes it possible 
to map the nuclear response as a function of its excitation energy with a 
spatial resolution that can be adjusted to the scale of processes that need to 
be studied.

Several decades of experimental and theoretical work on electron scattering 
have provided a wealth of information on the properties of stable nuclei. 
Nuclear charge density distributions and charge radii have been determined from
the analysis of elastic electron scattering data~\cite{DeJ:1987qc,Fricke:1995zz}.
At low energy loss bound nucleons may produce excited states as
a result of single-particle (s.p.) transitions and/or collective motion. 
Spectroscopic analysis of such excitation mechanisms is the traditional field 
of nuclear physics. The systematic investigation of inelastic electron 
scattering has given the necessary  support to the foundation of many-body 
theories applied to the nuclear systems. Spin, parities, and the strength and 
structure of the transition densities connecting the ground and excited states 
have been studied \cite{Donnelly:1975ze,Heisenberg:1984hb}.
In comparison with hadron scattering, which also gives important information, 
only electron scattering  can be used to determine the detailed spatial 
distributions of the charge transition densities for a variety of
single-particle and collective transitions.
At energy loss little above particle emission threshold the quasi-bound giant
resonances occur. Here the possibility of independently varying  $\omega$  and
{\boldmath${q}$}, in conjunction with the coincident detection of nucleons emitted in the
decay process, allows a detailed analysis of the various types of collective
motions that are responsible for contributions of different multipolarities to
such  resonances. Several electric and magnetic giant multipole resonances have
been discovered and studied using electron scattering \cite{Moya:1993,bookgdr}. 

At energies above giant resonances a large broad peak occurs at about 
$\omega=q^2/(2m_{N})$, where $m_{N}$ is the nucleon mass. Its 
position corresponds to the elastic peak in electron scattering by a free 
nucleon. It is quite natural to assume that a quasi-free process is responsible 
for such a peak with a nucleon emitted quasi-elastically.
If the nucleons were indeed free, the peak would be sharp and would
just occur at $\omega=q^2/(2m_{N})$, corresponding to the energy taken
by the recoiling free nucleon. A shift in the position of the peak is
produced by the nuclear binding, while a broadening of the peak is produced by
Fermi motion. 

Coincidence $(e,e^{\prime}p)$ experiments in the quasi-elastic (QE) region represent 
a very clean tool to explore the proton-hole 
states. A large amount of data for the exclusive $(e,e^{\prime}p)$ reaction 
have confirmed the assumption of a direct knockout mechanism and has provided 
accurate information on the s.p. structure of stable closed-shell 
nuclei 
\cite{Boffi:1993gs,book,Frullani:1984nn,Bernheim:1981si,Lapikas:1003zz,deWittHuberts:1990zy,Udias:1993xy,gao00}. 
The separation energy and the momentum distribution of the 
removed proton, which allows to determine the associated quantum numbers, have 
been obtained. From the comparison between experimental and theoretical 
cross sections it has been possible to extract the spectroscopic factors, 
which give a measurement of the occupation of the different shells and, as a 
consequence, of the effects of nuclear correlations, which go beyond a mean 
field description of nuclear structure. 

In the inclusive $(e,e')$ process only the scattered electron is detected and 
the final nuclear state is undetermined, but the main contribution in the 
region of the QE peak still comes from the interaction on 
single nucleons. In comparison with the exclusive $(e,e^{\prime}p)$ process, 
the inclusive $(e,e')$ scattering corresponds to an integral over all available
nuclear states and consequently provides less specific information, but it is
more directly related to the dynamics of the initial nuclear ground state. The
width of the QE peak can give a direct measurement of the average
momentum of nucleons in nuclei,  the shape depends on the
distribution in energy and momentum of the initially bound nucleons. Precise
measurements can give direct access to integrated properties of the nuclear
spectral function which describes this distribution. 
A considerably body of QE data for light-to-heavy nuclei in different
kinematic situations has been collected  
\cite{book,RevModPhys.80.189,web-benhar}. 
Not only differential cross sections,
but also the contribution of the separate longitudinal and transverse response
functions have been considered. From the theoretical point of view, many efforts 
have been devoted to the description of the available data and 
important progress has been achieved in terms of experimental results and
theoretical understanding \cite{Boffi:1993gs,book,RevModPhys.80.189}.   

The use of the electron probe can be extended to exotic nuclei. The detailed 
study of the properties of nuclei far from the stability line and the 
evolution of nuclear properties with respect to the asymmetry between the 
number of neutrons and protons is one of the major topics of interest in
 modern nuclear physics.

In the next years the advent of radioactive ion beams (RIB) facilities
\cite{tan95,gei95,mue01} will provide a large amount of data on unstable nuclei. 
A new generation of electron-RIB colliders using storage rings is under 
construction at RIKEN (Japan) \cite{sud01,sud09,kat03} and GSI (Germany) 
\cite{gsi06}. 
These facilities will offer unprecedented 
opportunities to study the structure of exotic unstable nuclei through electron 
scattering in the ELISe experiment at FAIR in 
Germany \cite{elise,Simon:2007zz,Antonov:2011zza} and the SCRIT project in 
Japan \cite{sud10,Suda01012012}. 
Kinematically complete experiments, where, in contrast to conventional 
electron scattering, all target-like reaction products are detected, will 
become feasible for the first time, allowing a clean separation of different
reaction channels as well as a reduction of the unavoidable radiative 
background seen in conventional experiments. Therefore, even applications using
stable isotope beams will be of interest.

Several papers devoted to theoretical treatments of electron scattering
off exotic nuclei have recently been published, underlining the usefulness of 
electron scattering for investigating the structure of unstable nuclei
\cite{Garrido:1999zy,Garrido:2000ht,Ershov:2005kq,PhysRevC.70.034303,Antonov04,
PhysRevC.72.044307,PhysRevC.75.024606,Khan:2007ji,PhysRevC.79.034318,
PhysRevC.79.044313,RocaMaza:2008cg,RocaMaza:2012hv,PhysRevC.79.014317,
PhysRevC.77.064302,Liu:2012zj,Dong:2012ed,Giusti:2011it,esotici1}.  
In this work we give another contribution to this field. Here we present and 
discuss numerical predictions for elastic and inclusive QE electron scattering 
cross sections on oxygen ($^{14-28}$O) and calcium ($^{36-56}$Ca) isotopic 
chains.  

The study of the evolution of nuclear properties along isotopic chains requires
a good knowledge of nuclear matter distributions for protons and neutrons 
separately. The ground state densities reflect the basic properties of effective
nuclear forces and provide fundamental nuclear structure information. 
Elastic electron scattering allows to measure with excellent precision only
charge densities and therefore proton distributions. It is much more difficult
to measure neutron distributions. Our present knowledge of neutron densities comes 
primarily from hadron scattering experiments, the analysis of which requires 
always model-dependent assumptions about strong nuclear forces at low-energies.
A model-independent probe of neutron densities is provided by
parity-violating elastic electron scattering, where  direct information on 
the neutron density can be obtained from the  measurement   
of the parity-violating asymmetry $A_{pv}$ parameter, which is defined as the 
difference between the cross sections for the scattering of right- and 
left-handed longitudinally polarized electrons 
\cite{Donnelly1989589,Donnelly19791,PhysRevC.57.3430}. This quantity is 
related to the radius of the neutron distribution $R_n$, because $Z^0$-boson 
exchange, which mediates the weak neutral interaction, couples mainly to 
neutrons and provides a robust model-independent measurement of $R_n$.

In 2012, the first measurement of $A_{pv}$ \cite{Abrahamyan:2012gp} 
(and weak charge form factor \cite{PhysRevC.85.032501}) in the elastic 
scattering of polarized electrons from $^{208}$Pb has been performed in
Hall A at the Jefferson Lab (experiment PREX).
The PREX collaboration obtained $A_{pv}$ = 0.656 $\pm$ 0.060(stat) 
$\pm$ 0.014(syst) ppm, corresponding to a difference between the radii of the 
neutron and proton distributions $R_n - R_p$ = 0.33 $^{+0.16} _{-0.18}$ fm. 
Unfortunately with such large uncertainties it is not possible to draw 
definite conclusions about the radius and the
distribution of neutrons in a heavy finite nucleus like $^{208}$Pb. 
The problems that affected the original setup will be 
strongly reduced by improving electronics and radiation protection (see
the recently approved PREX-II experiment \cite{prex2}). In addition, the
CREX experiment \cite{crex}, with the goal of measuring the neutron skin of 
$^{48}$Ca, has also been conditionally approved. The combined analysis of the 
PREX and CREX experiments will allow to test the assumptions of microscopic 
models by testing the dependence of $R_n$ on the atomic mass number $A$. 

From a theoretical point of view, parity violation in elastic electron 
scattering has been recently
studied by several authors within the framework of mean field approaches
\cite{Moreno2009306,0954-3899-39-1-015104,PhysRevLett.106.252501,
1742-6596-366-1-012011,Dong:2012ed,Liu:2012zj,PhysRevC.77.064302,
PhysRevC.79.014317} 
after almost a decade from the first detailed 
calculations \cite{PhysRevC.57.3430,PhysRevC.61.064307,PhysRevC.63.025501}. 
In particular, in Ref. \cite{PhysRevLett.106.252501} the authors suggested 
that a 1\% measurement of $A_{pv}$ can constrain the slope $L$ of the
symmetry energy close to a 10~MeV level of accuracy. 
A precise measurement of $R_n$ is definitely a crucial step to improve our 
knowledge of neutron-rich matter, i.e., the outer part of neutron stars. 
 
In this paper we present calculations of $A_{pv}$ performed for oxygen and 
calcium isotopic chains and test the isotopic dependence. In addition, we 
present a comparison with the  results of the PREX collaboration and provide
some estimates for the future experiment  CREX.

The basic ingredients of the calculations for both elastic and QE
scattering are the ground state wave functions 
of proton and neutron s.p. states,  for which we have used a 
relativistic mean field (RMF) model.
In the last years mean field approaches
have been very successfully employed in several aspects of 
nuclear structure phenomena. 
Effective hadron field theories with medium dependent parameterizations of the 
meson-nucleon vertices retain the basic
structure of the relativistic mean-field framework, 
but can be more directly related to the
underlying microscopic description of nuclear 
interactions \cite{Finelli:2003fk, Finelli:2004kd, Finelli:2005ni}.
After the first applications \cite{PhysRevC.52.3043,Typel1999331}, restricted 
to infinite systems and spherical nuclei,
with the parameterizations DD-ME1 \cite{PhysRevC.66.024306} and DD-ME2
\cite{PhysRevC.71.024312}, 
calculations have been extended to open shell nuclei, exotic systems,
superheavies, and collective resonances \cite{Vretenar2005101}.

The cross sections for elastic electron scattering are 
obtained from  the 
numerical solution of the partial wave Dirac equation and includes Coulomb 
distortion effects.

In inclusive QE electron scattering a proper description of the final-state
interactions (FSI) between the emitted nucleon and the residual nucleus is 
an essential ingredient for the comparison with data. For the calculations 
presented in this paper we have employed the relativistic Green's function 
model, which has been widely and successfully applied to the analysis of
QE electron and neutrino-nucleus scattering data on different 
nuclei
\cite{Meucci:2003uy,Meucci:2003cv,Meucci:2005pk,Meucci:2009nm,Meucci:2011pi,Meucci:ant,Meucci:2011nc,Meucci:2011vd}.  

As a first step, we test the results of the models in comparison with some of 
the data available for elastic and QE scattering on $^{16}$O and 
$^{40}$Ca nuclei. Then, with extensive calculations 
on oxygen and calcium isotopic chains, we investigate  
the evolution of some ground-state properties as a function of the neutron 
number.

The case of the exclusive quasi-free $(e,e^{\prime}p)$ reaction has been 
investigated in \cite{Giusti:2011it,Giusti:jp},  where the cross sections 
obtained with different relativistic and nonrelativistic approaches based on 
the mean-field description for the proton bound state wave function are 
compared for oxygen and calcium isotopic chains. 

The paper is organized as follows. In Sec. II we give the basic formalism
involved in the description of elastic, quasi-elastic, and parity-violating
electron scattering. In Sec. III we present a brief discussion 
of the relativistic mean-field model and of the calculations 
of the self-consistent ground-state proton and neutron densities 
of calcium and oxygen isotopes. In Sec. IV we outline the main features of the 
relativistic Green's function model, which is used 
to describe final-state interactions  in the inclusive quasi-elastic electron 
scattering.
In Sec. V we show and discuss our theoretical results obtained for elastic, 
quasi-elastic, and parity-violating electron scattering. Finally, in Sec. VI
we summarize our results and present our conclusions.


\section{Elastic and quasi-elastic electron scattering}
\label{sec.scattering}

\subsection{Elastic electron scattering}
\label{sec.elastic}

In the  one-photon exchange 
approximation and neglecting the effect of the nuclear Coulomb field on 
incoming and outgoing electrons, i.e. in the plane-wave Born approximation (PWBA), 
the differential cross section for the elastic scattering of an
lectron with momentum transfer $q$ off a spherical spin-zero nucleus  is given by
\inieq
\left(\frac{\diff \sigma}{ \diff \Omega^{\prime}}\right)_{{EL}} =
\sigma_{{M}} 
\abs{F_p(q)}^2
\ , \label{eq.csel}
\fineq
where  $\Omega^{\prime}$ is the scattered electron solid angle, 
$\sigma_{{M}}$ is the Mott cross section \cite{Boffi:1993gs,book}
and \inieq
F_p(q) = 
\int \diff \r ~ \jmath_0 (qr) \rho_p (r) \  ,   \label{eq.csfp}
\fineq
is the charge form factor for a spherical nuclear charge (point proton) density 
$\rho_p (r)$  and $\jmath_0$ is the
zeroth order spherical Bessel function.

The PWBA is, however, not adequate for medium and heavy nuclei where the
distortion produced on the electron wave functions by the nuclear Coulomb
potential $V (r)$ from $\rho_p (r)$ can have significant effects. 
The DWBA cross sections are obtained  from the numerical solutions of the
partial wave Dirac equation.

\subsection{Inclusive quasi-elastic electron scattering}
\label{sec.qe}

In the one-photon exchange approximation the inclusive differential cross 
section for the QE $(e,e^{\prime})$ scattering on a nucleus is obtained from the
contraction between the lepton and hadron tensors as  \cite{book} 
\inieq
\left(\frac{\diff \sigma}{\diff \varepsilon^{\prime} \diff
\Omega^{\prime}}\right)_{{QE}} =
\sigma_{M} 
\left[ v_L R_L + v_{T} R_{T}\right] 
\ , \label{eq.csqe}
\fineq
where  $\varepsilon^{\prime} $ is the energy of the scattered electron. 
The coefficients $v$ come from the components of the lepton tensor that, under
the assumption of the plane-wave approximation for the 
electron wave functions, depend only on the lepton kinematics,
\begin{eqnarray}
v^{}_{{L}} = \left(\frac {|Q^2|} {|\q|^2}\right)^2 \ , \ 
v^{}_{{T}}=\tan^2\frac {\theta}{2} - \frac{|Q^2|} {2|\q|^2} \  ,\ 
\label{eq.lepton}
\end{eqnarray}
where $\theta$ is the electron scattering angle and $Q^2 = |\q|^2 - \omega^2$. 
All nuclear structure information is contained in the longitudinal and 
transverse response functions $R_L$ and $R_T$, expressed by
\inieq
R_{{L}}(q,\omega) &=& W_{\textrm{}}^{00}(q,\omega) \ , \nonumber \\
R_{{T}}(q,\omega) &=& W_{\textrm{}}^{11}(q,\omega) + 
      W_{\textrm{}}^{22}(q,\omega) \  ,
\label{eq.response}
\fineq
in terms of the diagonal components of the hadron tensor, 
that is given by bilinear products of the transition matrix
elements of the nuclear electromagnetic many-body current operator $\hat{J}^{\mu}$ between
the initial state of the nucleus $\mid\Psi_0\rangle$, of energy $E_0$, and the 
final states $\mid \Psi_{ {f}} \rangle$, of energy $E_{ {f}}$, 
both eigenstates of the nuclear Hamiltonian $H$, as 
\begin{eqnarray}
 W^{\mu\mu}(q,\omega) &=& \overline {\sum_{ {i}}}
 \bint\sum_{ {f}}  \langle 
\Psi_{ {f}}\mid \hat{J}^{\mu}(\q) \mid \Psi_0\rangle \nonumber \\ &\times&
\langle 
\Psi_0\mid \hat{J}^{\mu\dagger}(\q) \mid \Psi_{ {f}}\rangle 
\ \delta (E_0 +\omega - E_{ {f}}),
\label{eq.ha1}
\end{eqnarray}
involving an average over the initial states and a sum over the undetected final 
states. The sum runs over the scattering states corresponding to all of the 
allowed asymptotic configurations and includes possible discrete 
states \cite{Boffi:1993gs,book}.

The hadron tensor can equivalently be expressed as
\begin{eqnarray}
 W^{\mu\mu}(q,\omega) &=& -\frac{1}{\pi}  \Im \langle \Psi_0 
 \mid J^{\mu\dagger}(\q)  G(E_{{f}}) J^{\mu}(\q) \mid \Psi_0 \rangle \ , 
\label{eq.ht1}
\end{eqnarray}
where $E_{{f}}=E_0 +\omega$ and $G(E_{{f}})$ is the many-body 
Green's function related to the many-body nuclear Hamiltonian $H$.

The hadron in tensor in Eq. (\ref{eq.ht1}) contains the full many-body propagator of
the nuclear system. As such, it is an extremely complicated object and some
approximations are needed to reduce the calculation of the nuclear response to 
a tractable form.

\subsection{Parity-violating electron scattering}
\label{sec.PVES}

When a photon is exchanged between two charged particles a $Z^0$ boson is also
exchanged. At the energies of interest in electron scattering the strength of
the weak process mediated by the $Z^0$ boson is negligible compared with the
electromagnetic strength. The role played by the $Z^0$ exchange is therefore not
significant unless an experiment is set up to measure a parity-violating
observable. While the electromagnetic interaction conserves parity, the weak
interaction does not and this is how we are sensitive to $Z^0$ exchange in
electron scattering.

The degree of parity violation can be measured by the parity-violating 
asymmetry $A_{pv}$, or helicity asymmetry, which is 
defined as the difference between the cross sections for the scattering of
electrons longitudinally polarized parallel  and antiparallel  
to their momentum.
This difference arises from the interference of photon and 
$Z^0$ exchange. As it has been shown in Refs. \cite{Donnelly1989589,PhysRevC.61.064307}, 
the asymmetry in the parity violating elastic polarized electron 
scattering represents an almost direct measurement of the 
Fourier transform of the neutron density.

The electron spinor for elastic scattering on a spin-zero nucleus
can be written as the solution of a Dirac equation with total potential
\begin{eqnarray}
U(r) = V(r) + \gamma_5 A(r) \ ,
\label{pot1}
\end{eqnarray}
where $V(r)$ is the Coulomb potential and $A(r)$ is the axial potential which
results from the weak neutral current amplitude and  which depends on the Fermi constant 
$G_F\simeq 1.16639 \times 10 ^{-11}$ MeV$^{-2}$, i.e., 
\begin{eqnarray}
A(r)= \dfrac{G_F}{2\sqrt{2}}\ \rho_W(r)\ ,
\label{pot2}
\end{eqnarray}
The weak charge density $\rho_W$ is related to the neutron density and it is defined 
\begin{eqnarray}
\rho_W(r)=  & \int & \diff \r^\prime \ G_E 
\left(|{\bf r}-{\bf r}^\prime|\right)  \nonumber \\
 & \times & \left[ -  \rho_n(r^\prime) +  
 (1-4{\sin}^2\Theta_W)\rho_p(r^\prime)\right] \ ,
\label{rhoW}
\end{eqnarray}
where $\rho_n$ and $\rho_p$ are point neutron and proton densities, 
$G_E(r)\approx \dfrac{\Lambda^3}{8\pi}\ e^{-\Lambda r}$ is the
electric form factor of the proton, with 
$\Lambda=4.27$ fm$^{-1}$, and
$\sin^2\Theta_W\simeq 0.23$ is the Weinberg angle. 
The axial potential of Eq. (\ref{pot2}) is much smaller than the vector potential and, since 
 $1-4{\sin}^2\Theta_W \ll  1$, it depends mainly on the
neutron distribution $\rho_n(r)$.
 

In the limit of vanishing electron mass, the helicity states 
$\Psi_\pm=\dfrac{1}{2}(1\pm\gamma_5)\Psi$
satisfy the Dirac equation 
\begin{equation}
[{\bm \alpha}\cdot {\bf p} + U_\pm(r)]\Psi_\pm = E \Psi_\pm \ ,
\end{equation}
with 
\begin{equation}
U_\pm(r) = V(r) \pm A(r)\ .
\end{equation}
The parity-violating asymmetry $A_{pv}$, or helicity asymmetry, is 
defined 
\begin{eqnarray}
A_{pv}=\dfrac{\dfrac{d\sigma_+}{d\Omega} - \dfrac{d\sigma_-}{d\Omega}}{
       \dfrac{d\sigma_+}{d\Omega} + \dfrac{d\sigma_-}{d\Omega}} \ ,
\label{AP}
\end{eqnarray}
where $+(-)$ refers to the elastic scattering on the potential $U_\pm(r)$.
In Born approximation,  neglecting strangeness contributions and the electric neutron 
form factor, the parity-violating asymmetry can be rewritten 
as \cite{PhysRevC.63.025501,1742-6596-312-9-092044} 
\begin{eqnarray}\label{AP2}
A_{pv} = \dfrac{G_F\ Q^2}{4 \sqrt{2}\ \pi \alpha} \left[
4 \sin^2 \Theta_W - 1 + 
\dfrac{F_n(q)}{F_p(q)} \right] \ .
\end{eqnarray}
Since $4 \sin^2 \Theta_W - 1$ is small and $F_p(q)$ is known, we see that  
$A_{pv}$ provides a practical method to measure
the neutron form factor $F_n(q)$ and hence the neutron radius.
For these reasons parity-violating electron
scattering (PVES) has been suggested 
as a clean and powerful tool for
measuring the spatial distribution of neutrons in nuclei.

In QE electron scattering the helicity asymmetry is obtained in terms of
kinematic coefficients $v$ from the lepton tensor and of nuclear response functions 
as \cite{Meucci:2005pk}
\begin{eqnarray}
A^{{QE}}_{pv} =  A_0 \frac {v^{}_{{L}}R_{{L}}^{{AV}} +
 v^{}_{{T}}
R_{{T}}^{{AV}} + v_{{T}}'R_{{T}}^{{VA}}} 
{v^{}_{{L}}R_{{L}} + v^{}_{{T}}R_{{T}}}\ .
\label{eq.A}
\end{eqnarray}
The factor $A_0$ is defined as
\begin{equation}
A_0 = \dfrac{G_F\ Q^2}{2\sqrt{2}\pi\alpha} \ , \label{eq.a0}
\end{equation} 
where 
$\alpha$ is the fine structure constant. The denominator 
in Eq.~(\ref{eq.A}) contains the parity-conserving cross 
section of Eq.~(\ref{eq.csqe}), the numerator the parity-violating 
contribution, 
\begin{equation}
v'_{{T}}= \tan \frac {\theta}{2} 
\sqrt{\tan^2\frac {\theta}{2} +
 \frac{Q^2} {|\q|^2} }
\label{eq.leptonpv}
\end{equation}
and the response functions $R$ are given in terms of the polarized components 
of the hadron tensor $W_{{I}}^{\mu\nu}$ \cite{Meucci:2005pk} as
\begin{eqnarray} 
R_{{L}}^{{AV}} &=& g_{ A}W_{{I}}^{00}\ , \ 
R_{{T}}^{{AV}} = g_{ A}\left(W_{{I}}^{11}
+ W_{{I}}^{22}\right)\ , \nonumber \\
R_{{T}}^{{VA}} &=& i g_{ V}\left(W_{{I}}^{12} -  
W_{{I}}^{21}\right)\ ,
\label{eq.rf}
\end{eqnarray}
where the superscript AV denotes interference of axial-vector leptonic current
with vector hadronic current (the reverse for VA) and 
the couplings $g_{ A} = -\dfrac{1}{2}$ and $g_{ V} = -\dfrac{1}{2} + 2
\sin^2{\Theta_{ W}} \simeq -0.04$. 


\section{Relativistic model for ground-state observables}
\label{rmf}

\begin{figure}[tb]
\begin{center}
\vskip -.8cm
\includegraphics[scale=0.35]{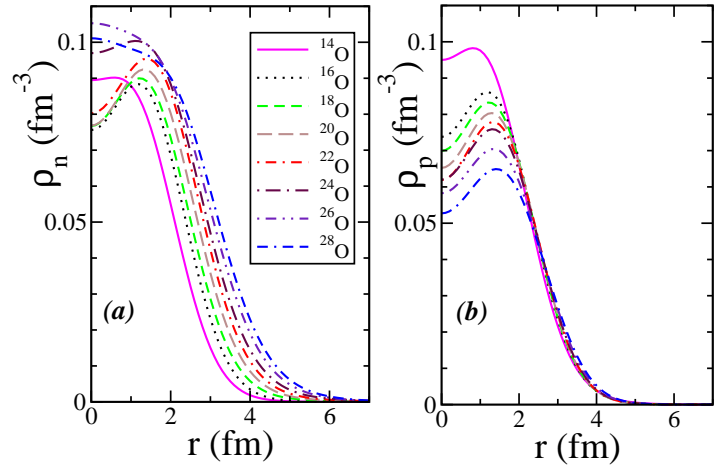} 
\end{center}
	\caption{(Color online)  Neutron, panel (a), and proton, panel 
	(b), distributions for the various oxygen isotopes we have considered.
\label{fig_den_O}}

	\end{figure}

\begin{figure}[tb]
\begin{center}
\vskip -.8cm
\includegraphics[scale=0.35]{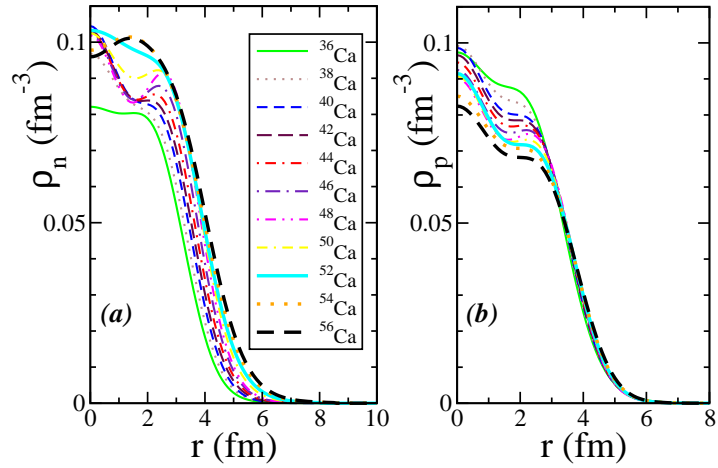} 
\end{center}
	\caption{(Color online)  The same as in Fig.~\ref{fig_den_O}, but for calcium isotopes.
\label{fig_den_Ca}}
	\end{figure}

In the standard representation of relativistic mean field approaches the 
nucleus is described as a system of Dirac
nucleons coupled to the exchange mesons and the electromagnetic field through an effective Lagrangian.
The isoscalar scalar-meson ($\sigma$), the isoscalar vector-meson ($\omega$), and the 
isovector vector-meson ($\rho$) build the
minimal set of meson fields that together with the electromagnetic field ($\gamma$)
is necessary for a quantitative description of bulk and s.p. nuclear properties.
The model is defined by the Lagrangian density
\begin{equation}
\mathcal{L} = \mathcal{L}_N+ \mathcal{L}_m + \mathcal{L}_{int} \; ,
\end{equation}
where $\mathcal{L}_N$ denotes the Lagrangian of the free nucleon, $\mathcal{L}_m$
is the Lagrangian of the free meson fields and the simplest set of interaction terms is 
contained in $\mathcal{L}_{int}$:
\begin{equation}
\mathcal{L}_{int} = - g_\sigma \bar{\psi}\sigma \psi - g_\omega \bar{\psi} \gamma^\mu \omega_{\mu} \psi -
g_\rho \bar{\psi} \gamma^\mu \vec{\tau} \cdot \vec{\rho}_\mu \psi \; .
\end{equation}
The couplings of the $\sigma$-meson and $\omega$-meson to the nucleon are
assumed to be of the form:
\begin{equation}
 g_{i}(\rho) = g_{i}(\rho_{sat})f_{i}(x) \quad\text{for} \quad i=\sigma
,\omega\;, \label{gcoupl}%
\end{equation}
where
\begin{equation}
f_{i}(x) = a_{i}\frac{1+b_{i}(x+d_{i})^{2}}{1+c_{i}(x+d_{i})^{2}}
\label{fcoupl}%
\end{equation}
is a function of $x=\rho/\rho_{sat}$ and $\rho_{sat}= 0.152$  fm$^{-3}$ 
denotes the nucleon
density at saturation in symmetric nuclear matter. Constraints at nuclear
matter saturation density and at zero density are used to reduce the number
of independent parameters in Eq.~(\ref{fcoupl}) to three. Three additional
parameters in the isoscalar channel are $g_{\sigma}(\rho_{sat}),\;
g_{\omega}(\rho_{sat})$, and $m_{\sigma}$, that is the mass of the 
phenomenological $\sigma$ meson. For the $\rho$ meson coupling the functional 
form of the
density dependence is suggested by Dirac-Brueckner calculations of asymmetric
nuclear matter:
\begin{equation}
g_{\rho}(\rho)=g_{\rho}(\rho_{sat})\exp[-a_{\rho}(x-1)]\;,
\end{equation}
and the isovector channel is parametrized by $g_{\rho}(\rho_{sat})$ and
$a_{\rho}$. Bare values are used for the masses of the $\omega$ and $\rho$
mesons: $m_{\omega}=783$ MeV and $m_{\rho}=763$ MeV.
DD-ME2 is determined by eight independent parameters, 
adjusted to the properties of symmetric
and asymmetric nuclear matter, binding energies, charge radii, and neutron
radii of spherical nuclei~\cite{PhysRevC.71.024312}. The interaction has been
tested in the calculation of ground state properties of a large set of spherical
and deformed nuclei. When
used in the relativistic random-phase approximation, DD-ME2 reproduces with high accuracy data on
isoscalar and isovector collective excitations.
For open-shell nuclei we employed a schematic ansatz: the constant gap approximation with
empirical $\Delta$ given by the 5-point formula \cite{Moller:1992zz}
\begin{eqnarray}
\label{delta5}
\Delta^{(5)} (N_0) = -\frac{1}{8}  & \bigg[ & E(N_0+2) - 4E(N_0+1) + 6E(N_0)
\nonumber \\
&- &  4E(N_0-1) + E(N_0-2) \bigg] \;.
\end{eqnarray}
In Figs.~\ref{fig_den_O} and \ref{fig_den_Ca} we plot the neutron (proton) 
density distributions $\rho_{n(p)}$ as a function of the radial coordinate $r$ 
for oxygen and calcium isotopes, respectively. These density distributions are the sum of 
the squared moduli  of the s.p. neutron (proton) wave functions.
All the nuclei we have investigated resulted to be bound.  
From the experimental point of view it seems rather well established that the 
neutron drip line for the oxygen isotopes starts with 
$^{26}$O \cite{Schiller:2005rp} and, therefore, $^{28}$O should not be bound.

For both isotopic chains the differences between the proton and neutron 
densities in Figs.~\ref{fig_den_O} and \ref{fig_den_Ca} generally increase 
with the neutron number. When the number of 
neutrons increases there is a gradual increase of the neutron radius. The 
differences of the neutron density
profiles in the nuclear interior display pronounced shell effects. 
The effect of adding neutrons is to populate and extend the neutron densities
and, to a minor extent, also the proton densities. 
In the case of protons, however, there is a decrease of the density in the 
nuclear interior to preserve the normalization to the constant number of
protons.      


\section{Relativistic Green's function model for inclusive quasi-elastic electron 
scattering\label{rgfs}}

In the QE region the nuclear response is dominated by
one-nucleon knockout processes, where the scattering occurs
with only one nucleon that is subsequently emitted. The
remaining nucleons of the target behave as simple spectators and QE electron 
scattering can adequately be described in
the relativistic impulse approximation (RIA) by the sum of
incoherent processes involving only one nucleon scattering and the components 
of the hadron tensor of Eq.~(\ref{eq.ha1}) are obtained from the 
sum, over all the s.p. shell-model states, of the squared absolute value of the transition 
matrix elements of the single-nucleon current.

A reliable description of final-state interactions (FSI) between the ejected nucleon
and the residual nucleus is an essential ingredient for the
comparison with data.
In the case of exclusive $(e,e^{\prime}p)$ processes,
 the use of complex optical potentials in the relativistic distorted
wave impulse approximation (RDWIA) has been able to successfully
describe a wide number of experimental data~\cite{Udias:1993xy,Meucci:2001qc,Meucci:2001ja,Meucci:2001ty,
Radici:2003zz,Tamae:2009zz,Giusti:2011it,esotici1}. 
It is clear that the pure RDWIA approach, based on the use of an absorptive complex potential,
would be inconsistent in the analysis of inclusive scattering,
where all final-state channels should be retained
and the total flux, although redistributed among all possible
channels due to FSI, must be conserved. 
Different approaches have been used to describe FSI in RIA calculations for 
the inclusive QE electron- and neutrino-nucleus
scattering~\cite{Maieron:2003df,Meucci:2003uy,Meucci:2003cv,
Meucci:2004ip,Meucci:2005pk,
Meucci:2006cx,Meucci:2006ir,Caballero:2006wi,
Meucci:2008zz,
Meucci:2009nm,Giusti:2009sy,Caballero:2009sn,Butkevich:2010cr,
Meucci:2011pi,Meucci:2011vd,but12,Meucci:ant}. 
In the relativistic plane-wave impulse approximation (RPWIA), FSI 
are simply neglected. In another approach, FSI are included in  calculations 
where the final nucleon state is evaluated with real potentials, either 
retaining only the real part of the relativistic energy-dependent complex 
optical potential (rROP) \cite{Butkevich:2010cr,but12}, or 
using the same relativistic mean field potential considered in 
describing the initial nucleon state \cite{Caballero:2006wi}.

In the relativistic Green's function (RGF) model FSI 
are described  in the inclusive process consistently with the exclusive 
scattering by the same complex optical potential, but the imaginary part is 
used in the two cases in a different way and in the inclusive scattering it 
redistributes the flux in all the channels and the total flux is conserved. 
Detailed discussions of the RGF model can be found in Refs. \cite{Capuzzi:1991qd,Meucci:2003uy,
Meucci:2003cv,Capuzzi:2004au,Meucci:2005pk,
Meucci:2009nm,Meucci:2011vd,Meucci:2011pi,Meucci:2011nc,Meucci:ant}.  
The model assumes that the ground state 
of the nucleus $|\Psi_0\rangle   $  is  non-degenerate, this is a suitable 
approximation for the even isotopes of oxygen and calcium considered in this
work,  with spin and parity $0^+$ \cite{bnlw}.

In the RGF model with suitable approximations, which are mainly
related to the impulse approximation, the components of the nuclear response
of Eq. (\ref{eq.ht1}) are written in terms of the s.p. optical model Green's
function. 
The spectral representation of the s.p. Green's function, which is
based on a biorthogonal expansion in terms of a non-Hermitian
optical potential and of its Hermitian conjugate, can be exploited to avoid the
explicit calculation of the s.p. Green's function and obtain the 
components of the hadron tensor in the form \cite{Meucci:2003uy}
\inieq
&  & W^{\mu\mu} (  q,  \omega )  =  \sum_n \Bigg[ \Re \ T_n^{\mu\mu}
(E_{{f}}-\varepsilon_n, E_{{f}}-\varepsilon_n)  \nonumber \\ 
& - & \frac{1}{\pi} \mathcal{P}  \int_M^{\infty} \diff \mathcal{E} 
\frac{1}{E_{{f}}-\varepsilon_n-\mathcal{E}} 
\ \Im \ T_n^{\mu\mu}
(\mathcal{E},E_{{f}}-\varepsilon_n) \Bigg] \ , \label{eq.finale}
\fineq
where $\mathcal{P}$ denotes the principal value of the integral, $n$ is the 
eigenstate of the residual nucleus with energy 
$\varepsilon_n$, and
\inieq
T_n^{\mu\mu}(\mathcal{E} ,E) &=& \lambda_n  \langle \varphi_n
\mid j^{\mu\dagger}(\q) \sqrt{1-\mcv'(E)}
\mid\tilde{\chi}_{\mathcal{E}}^{(-)}(E)\rangle \nonumber \\
&~ & \times  \langle\chi_{\mathcal{E}}^{(-)}(E)\mid  \sqrt{1-\mcv'(E)} j^{\mu}
(\q)\mid \varphi_n \rangle  \ . \label{eq.tprac}
\fineq
The factor $\sqrt{1-\mcv'(E)}$, where $\mcv'(E)$ is the energy derivative of 
the optical potential, accounts for interference effects between different 
channels and justifies the replacement in the calculations of the Feshbach 
optical potential $\mcv$ of the RGF model by the local phenomenological optical 
potential~\cite{Meucci:2003uy,Capuzzi:1991qd,Capuzzi:2004au}. 

Disregarding the square root correction, the second matrix element in 
Eq.~(\ref{eq.tprac}) is the transition amplitude of the usual RDWIA model for 
the exclusive single-nucleon knockout. In this matrix element
$j^{\mu}$  is the one-body nuclear current, $\chi^{(-)}$ is the 
s.p. scattering state of the emitted nucleon with energy $\mathcal{E}$,   
$\varphi_n$ is the overlap between the ground state of the target and the final state $n$, 
i.e., a s.p. bound state, and the spectroscopic factor $\lambda_n$ is the norm
of the overlap function.  In the model  
$\varphi_n$ and $\chi^{(-)}$  are consistenlty derived as eigenfunctions of 
the energy-dependent optical-model Hamiltonian at bound and scattering energies. 

In the exclusive one-nucleon knockout the imaginary part of the optical 
potential accounts for the flux lost in the 
channel $n$  towards the channels different from $n$, which are not included in
the exclusive process. In the inclusive response, where all the channels are
included, this loss is compensated by a corresponding 
gain of flux due to the flux lost, towards the channel $n$, in the other final 
states asymptotically originated by the channels different from $n$. 
This compensation is performed by the first matrix element in the right hand 
side of Eq.~(\ref{eq.tprac}),  which involves the eigenfunction 
$\tilde{\chi}_{\mathcal{E}}^{(-)}(E)$ of the Hermitian conjugate optical
potential, where the imaginary part has an opposite sign and has the 
effect of increasing the strength. Therefore, in the RGF approach the 
imaginary part of the optical
potential redistributes the flux lost in a channel in the other channels, and 
in the sum over $n$ the total flux is conserved.  
If the imaginary part of the
optical potential is neglected, the second term in Eq.~(\ref{eq.finale}) 
vanishes and, but for the square root factor, the first term gives the rROP 
approach.

In Eq. (\ref{eq.tprac}) $\tilde\chi^{(-)}$ and $\chi^{(-)}$ are therefore
eigenfunctions of the optical 
potential $\mcv(E)$ and of its Hermitian conjugate $\mcv^{\dagger}(E)$ , which 
are nonlocal operators with a possibly complicated matrix structure. Neither
microscopic nor empirical calculations of $\mcv(E)$ are 
available. Only phenomenological local optical potentials, obtained through 
fits to  elastic nucleon-nucleus scattering data, are available. 
These phenomenological optical potentials are used in RGF calculations. 
As no relativistic optical potentials are available for the bound states, the 
overlap functions $\varphi_n$, are computed in the present work using the 
model discussed in Sect. \ref{rmf}. 

The RGF model has been applied to parity-violating QE electron scattering in
Ref. \cite{Meucci:2005pk}. The main steps of the model are the same, the
expressions for the electromagnetic-weak interference components of the hadron
tensor $W_{{I}}^{\mu\nu}$ can be found in Ref. \cite{Meucci:2005pk}.

\section{results}
\label{results}

In this section we present and discuss numerical predictions 
for elastic and QE electron scattering which can hopefully be useful 
for future measurements in experimental RIB facilities.
We study the evolution of some electron scattering observables in
isotopic chains of medium  systems, which are exemplified by the cases of the 
oxygen and calcium isotopes.
Many of these nuclei lie in the region of the nuclear chart that
is likely to be explored  in future electron-scattering experiments.
As a first step, with a few numerical examples we test the results of our 
models in comparison with available data. Then, for each nucleus in an isotopic 
chain, we compute and compare the associated elastic and QE cross sections and
parity-violating asymmetry in order to obtain information on the effects of 
isospin asymmetry on nuclear structure.  

\subsection{Elastic electron scattering}
\label{res_elastic}

The cross sections for elastic electron scattering have been calculated in the
DWBA and with the self-consistent relativistic ground state charge densities 
described in Sect. \ref{rmf}. In the PWBA the cross section is proportional to 
the Fourier transform of the proton charge density (see Eq.~(\ref{eq.csfp})) 
and reflects its behavior
also when Coulomb distortion is included in the calculations. 
In different studies of the charge form factors along isotopic chains 
\cite{PhysRevC.72.044307,PhysRevC.76.044322,PhysRevC.71.054323,RocaMaza:2008cg,PhysRevC.79.044313} 
it has been found that, when the number of neutrons increases, the squared 
modulus of the charge form factor and the position of its minima show, 
respectively, an upward trend and a significant inward  shifting in the 
momentum transfer.  

\begin{figure}[tb]
\begin{center}
\includegraphics[scale =0.45]{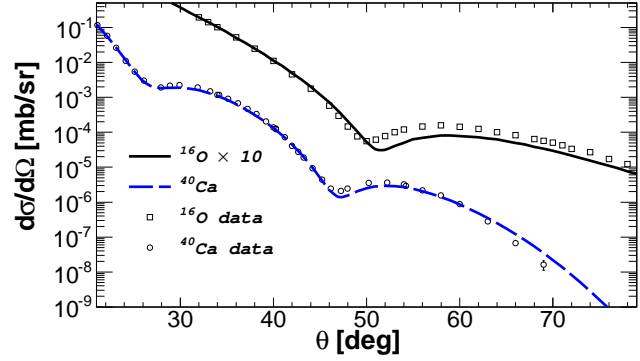} 
\end{center}
	\caption{(Color online)  Differential cross section for elastic 
	electron scattering on $^{16}$O at an electron energy $\varepsilon =
	374.5$ MeV and $^{40}$Ca at $\varepsilon = 496.8$ MeV as a function of the 
	scattering angle $\theta$. Experimental data from
	\cite{PhysRevC.7.1930} ($^{16}$O) and \cite{Sick:1970ma} ($^{40}$Ca).
\label{fig_elastic-exp}}
	\end{figure}

An example of the comparison between theoretical and experimental differential 
cross sections is displayed in 
Fig.~\ref{fig_elastic-exp} for elastic electron scattering  
on $^{16}$O at an electron energy $\varepsilon = 374.5$ MeV and on $^{40}$Ca at 
$\varepsilon = 496.8$ MeV.   
The general trend of the experimental data is reasonably reproduced by
the calculations. Both experimental cross sections considered in the figure are 
well described  at low scattering angles. For $^{40}$Ca there is a fair 
agreement between theory and data also at larger angles, while for $^{16}$O data beyond the minimum 
are somewhat underestimated by the theoretical results.

\begin{figure}[tb]
\begin{center}
\includegraphics[scale=0.45]{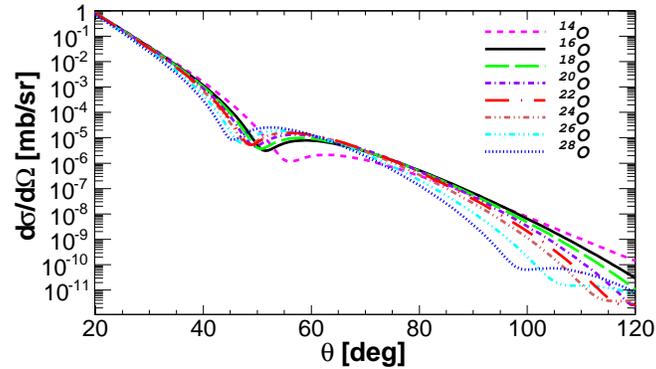} 
\end{center}
	\caption{(Color online)  Differential cross section for elastic 
	electron scattering on $^{14-28 }$O at  $\varepsilon = 374.5$ MeV
	 as a function of  $\theta$.
\label{fig_elastic-o}}
	\end{figure}

\begin{figure}[tb]
\begin{center}
\includegraphics[scale=0.45]{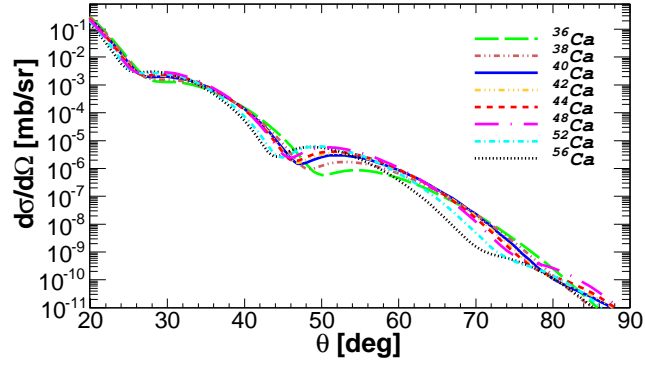} 
\end{center}
	\caption{(Color online)  Differential cross section for elastic 
	electron scattering on $^{36-56 }$Ca at  $\varepsilon = 496.8$ MeV
	 as a function of  $\theta$.
\label{fig_elastic-ca}}
	\end{figure}

The calculated differential cross sections for elastic electron scattering on
various oxygen isotopes ($^{14-28}$O) at $\varepsilon = 374.5$ MeV and on 
calcium isotopes ($^{36-56}$Ca) at $\varepsilon = 496.8.5$ MeV are shown in 
Figs.~\ref{fig_elastic-o} and \ref{fig_elastic-ca}, respectively.
With increasing neutron number the positions of the diffraction minima shift 
toward  smaller scattering angles, i.e., towards smaller values of the momentum transfer.
The shift of the minima towards smaller $q$ is in general accompanied by a
simultaneous increase in the height of the maxima.
The behavior is similar for both isotopic chains here considered and is in
agreement with the results found in previous studies of charge form factors on
various isotopic chains, which were carried out with different mean-field models
\cite{PhysRevC.72.044307,PhysRevC.76.044322,PhysRevC.71.054323,RocaMaza:2008cg,PhysRevC.79.044313}. 

\subsection{Quasi-elastic electron scattering}
\label{res_qe}

The cross sections for QE electron scattering have been computed with the RGF
model discussed in Sect. \ref{rgfs}. 
Some results obtained in the RPWIA are also presented for a comparison.
In the calculations of the matrix elements in Eq.~(\ref{eq.tprac}) the
s.p. bound nucleon states are obtained from the relativistic
mean-field model with density-dependent meson-nucleon vertices and the 
DD-ME2 parametrization as described in Sect. \ref{rmf}. The s.p. scattering states
are eigenfunctions of the energy-dependent and  $A-$dependent ($A$ is the mass 
number) parameterization for the relativistic optical potential of 
Ref. \cite{Cooper:2009}, which is fitted to proton elastic
scattering data on several nuclei in an energy range up to 1040 MeV. 
The different number of neutrons along the O and Ca 
isotopic chains produces different optical potentials 
(see Ref. \cite{Cooper:2009} for more details).
For the
single-nucleon current we have used the relativistic free nucleon expression
denoted as CC2 \cite{DeForestJr1983232,Meucci:2003uy}.

The predictions of the RGF model have been compared with experimental data for QE
electron- and neutrino-nucleus scattering in a series of papers 
\cite{Meucci:2003uy,Meucci:ant,Meucci:2011nc,Meucci:2011vd,Meucci:2011pi,
Giusti:2009sy,Meucci:2009nm,Meucci:2006cx,Meucci:2005pk,Meucci:2003cv},
 where the calculations have been performed with different relativistic 
 mean-field models for the bound states and different parameterizations of the 
 relativistic optical potential.
   
\begin{figure}[tb]
\begin{center}
\includegraphics[scale=0.45]{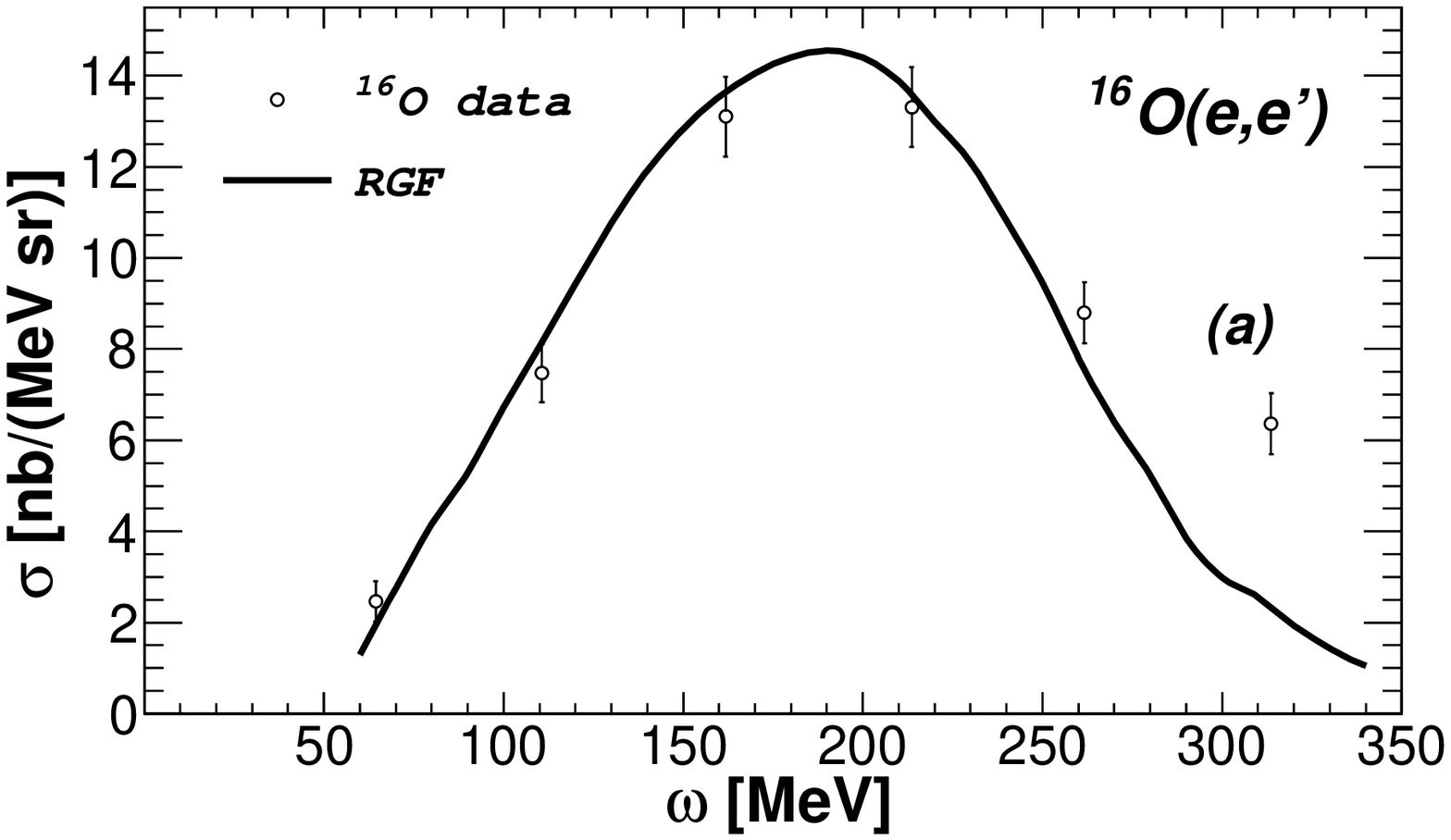} 
\includegraphics[scale=0.45]{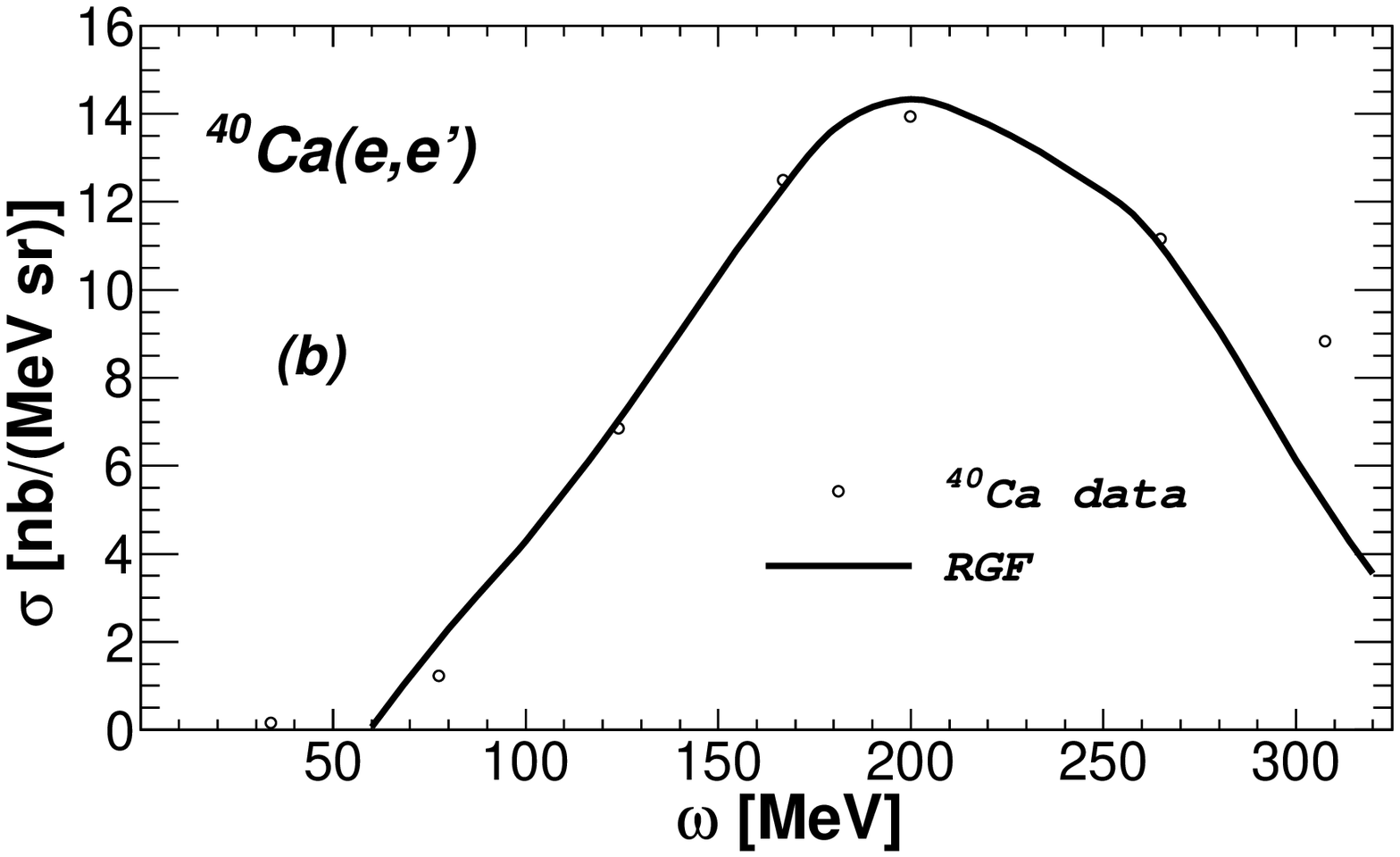} 
\end{center}
	\caption{Differential cross section of the reactions
	$^{16}$O$(e,e^{\prime})$, panel (a), and $^{40}$Ca$(e,e^{\prime})$, 
	panel (b), for different beam energies and electron scattering angles, 
$\varepsilon = 1080$ MeV and $\theta = 32^{\mathrm{o}}$ for
$^{16}$O$(e,e^{\prime})$ and $\varepsilon = 841$ MeV and 
$\theta = 45.5^{\mathrm{o}}$ for $^{40}$Ca$(e,e^{\prime})$,	
as a function of the energy transfer $\omega$. The RGF results are compared with  
the experimental data from \cite{Anghinolfi:1996vm} ($^{16}$O$(e,e^{\prime})$) 
and \cite{Williamson:1997} ($^{40}$Ca$(e,e^{\prime})$).
\label{fexpgf}}
	\end{figure}

In Fig.~\ref{fexpgf} our  RGF  results are compared with
the experimental $(e,e^{\prime})$ cross sections for two different kinematics 
on $^{16}$O and $^{40}$Ca target nuclei \cite{Anghinolfi:1996vm,Williamson:1997}. 
The agreement with the data is satisfactory, at least in the energy region of 
the QE peak.  The RGF model was developed to describe FSI in inclusive QE
electron scattering and is in general able to give a reasonable and even good
description of QE data. For energy regions below and above the QE peak other 
contributions, not included in the RGF model, can be important. Even in the QE
region, the relevance of contributions like meson exchange currents and Delta 
effects should be carefully evaluated before definite conclusions can be drawn
about the comparison with data
\cite{PhysRevC.69.035502,PhysRevC.71.015501,PhysRevC.77.034612}. 
Such contributions may be significant even in 
the QE region, in particular in kinematics where the transverse component of 
the nuclear response plays a major role in the cross section.

\begin{figure}[tb]
\begin{center}
\includegraphics[scale=0.45]{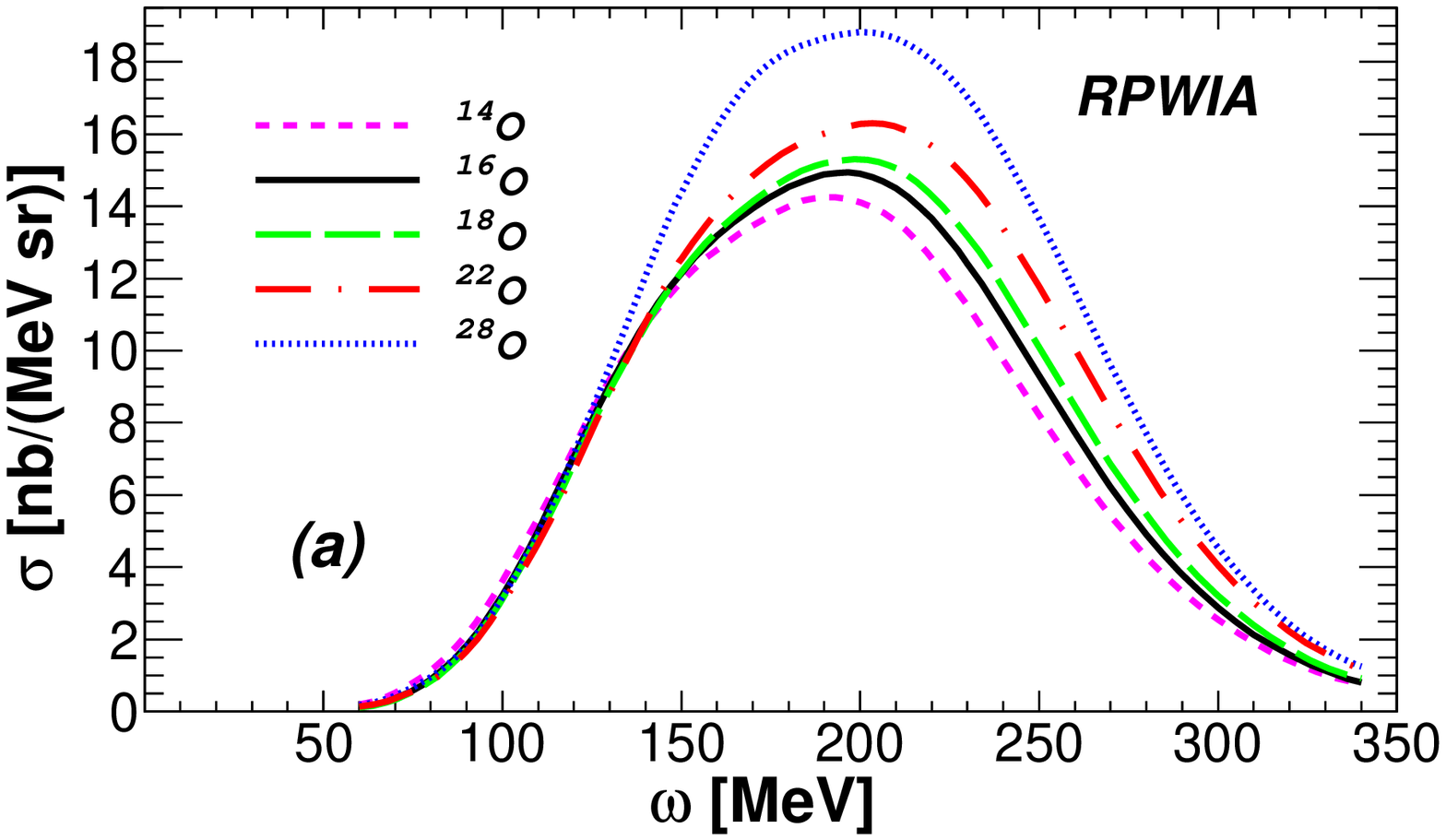} 
\includegraphics[scale=0.45]{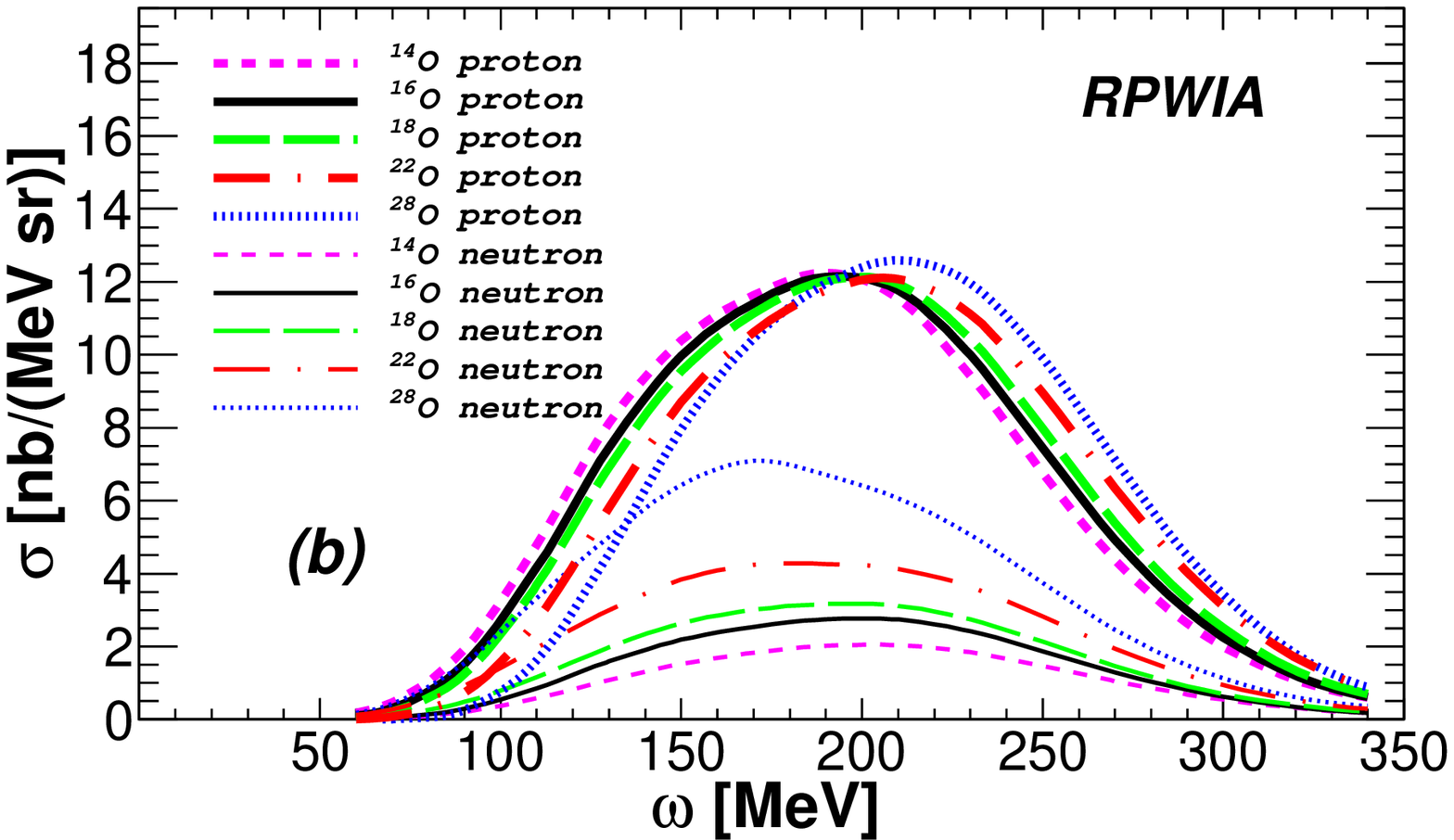} 
\end{center}
	\caption{(Color online) In panel (a) the differential RPWIA cross section for the 
	inclusive QE 
	$(e,e^{\prime})$  reaction on $^{14-28}$O at $\varepsilon = 1080$ MeV
	 and $\theta = 32^{\mathrm{o}}$ is shown as a function of $\omega$. 
	In panel (b) the separate contributions of protons (thick lines) and 
	neutrons (thin lines) are displayed. 
	\label{foiso-pw}}
	\end{figure}

The cross section of the inclusive QE 
	$(e,e^{\prime})$  reaction on $^{14-28}$O isotopes at 
	$\varepsilon = 1080$ MeV and $\theta = 32^{\mathrm{o}}$ are 
	shown  in Fig.~\ref{foiso-pw}.
In a first approximation, we have neglected FSI and calculations have been 
performed in the RPWIA. In this case, the differences between the results 
for the various isotopes are entirely due to the differences in the s.p. bound
state wave functions of each isotope.  While only the charge proton density
distribution contributes to the cross section of elastic electron scattering, 
the cross section of QE electron scattering is obtained from the 
sum of all the integrated exclusive one-nucleon knockout processes, due to the 
interaction of the probe with all the individual nucleons, protons and 
neutrons, of the nucleus and contains information on the dynamics 
of the initial nuclear ground state. The separate contributions from 
protons and neutrons are also shown in the lower panel of Fig.~\ref{foiso-pw}.
In an usual experiment where only the scattered electron is detected these two
quantities cannot be separated experimentally, but their comparison can give
useful information on the different role of  protons and neutrons in the
inclusive QE cross section. The main role is played by protons, which give most
of the  contribution.
Increasing the neutron number it is quite natural to understand the proportional
increase of the neutron contribution. No significant increase is found in the proton
contribution. Thus, the increase of the cross section in the upper panel of 
the figure is due to the increase of the neutron contribution.
The shift of the proton contribution towards higher values of $\omega$ seen 
in the figure is mainly related to the increase of the proton separation 
energy with increasing neutron number (increasing the neutron number 
the protons experience more binding and their separation energies increase) 
than to changes in the proton wave functions.  
A different and opposite shift can be seen in the case of the neutron contribution 
and, therefore, the final effect is that the shift is strongly reduced in the 
QE cross section shown in the upper panel of Fig.~\ref{foiso-pw}.

\begin{figure}[tb]
\begin{center}
\includegraphics[scale=0.45]{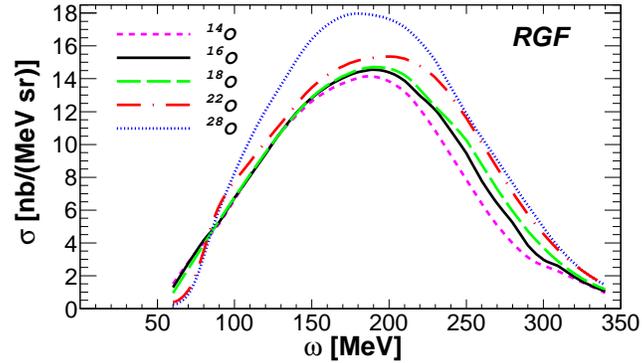} 
\end{center}
	\caption{(Color online) Differential RGF cross section for the inclusive QE
	$(e,e^{\prime})$  reaction on $^{14-28}$O in the same kinematics as in
	Fig.~\ref{foiso-pw}.
	\label{foiso-gf}}
	\end{figure}

 In Fig.~\ref{foiso-gf} we show the QE $(e,e^{\prime})$ cross sections
 calculated for oxygen isotopes with the RGF model and in the same kinematics 
 as in Fig.~\ref{foiso-pw}. 
The general trend  of the cross sections, their magnitude, and their evolution 
with respect to the change of the neutron number are generally similar in 
RPWIA and RGF. 
The FSI effects in the RGF calculations produce, however, some differences 
which can be seen in the low energy transferred region, 
where the cross sections for $^{14,16,18}$O are enhanced with respect to those 
for $^{22}$O and $^{28}$O. In addition, the shift towards higher $\omega$ is 
more significant than in the RPWIA case, but for $^{28}$O. 

\begin{figure}[tb]
\begin{center}
\includegraphics[scale=0.45]{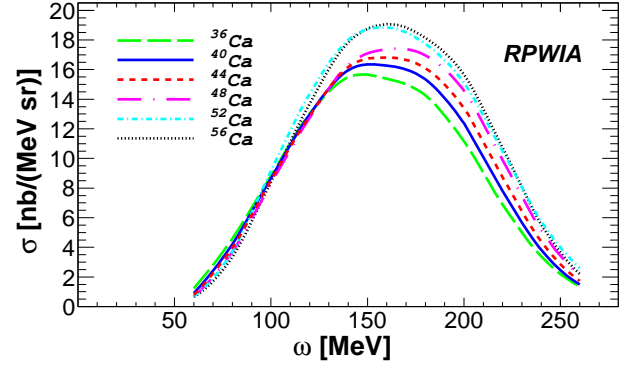} 
\end{center}
	\caption{(Color online) Differential RPWIA cross section for the inclusive QE 
	$(e,e^{\prime})$  reaction on $^{36-56}$Ca at $\varepsilon = 560$ MeV
	 and $\theta = 60^{\mathrm{o}}$ as a function of $\omega$. 
	\label{fcaiso-pw}}
	\end{figure}

\begin{figure}[tb]
\begin{center}
\includegraphics[scale=0.45]{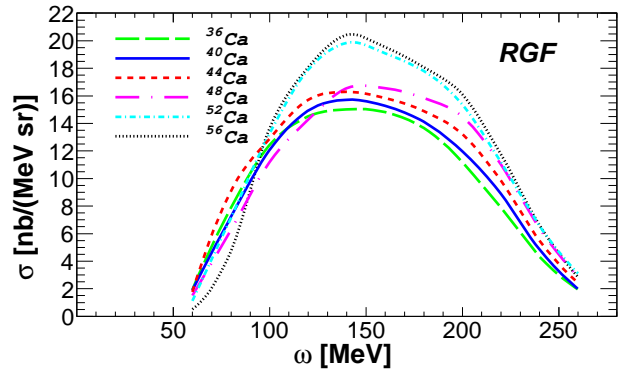} 
\end{center}
	\caption{(Color online) The same as in Fig.~\ref{fcaiso-pw}, but in the RGF model.
	\label{fcaiso-gf}}
	\end{figure}

The cross section of the inclusive QE $(e,e^{\prime})$  reaction on 
$^{36-56}$Ca isotopes at 
	$\varepsilon = 560$ MeV and $\theta = 60^{\mathrm{o}}$ calculated in
	the RPWIA and in the RGF are 
	shown  in  Figs.~\ref{fcaiso-pw} and \ref{fcaiso-gf}, respectively.
The general behavior of the cross sections and their evolution with
increasing neutron number is similar for calcium  and oxygen isotopes. 
 The magnitude increases with the neutron number, but FSI effects are somewhat 
 more visible for calcium isotopes. The RGF cross sections on 
 $^{36,40,44,48}$Ca in Fig.~\ref{fcaiso-gf} 
 are enlarged over a wide range of $\omega$ and are slightly reduced with 
 respect to the RPWIA results in Fig.~\ref{fcaiso-pw}. This is particularly 
 visible for $^{48}$Ca and produces an apparently large gap between the 
 cross sections of $^{48}$Ca and $^{52}$Ca. 
 
 As a final comment, we can add that interesting and peculiar effects are obtained
 in the evolution of QE inclusive cross section along isotopic chains, 
 but it is not easy to relate them to changes in the 
 matter distribution, which can be significant, particularly in the center of 
 the nucleus.

\subsection{Parity-violating asymmetry}
\label{parity}

The calculation starts with the self-consistent relativistic 
ground state proton and neutron densities (see Sect. \ref{rmf}).
The charge and weak densities are calculated by folding 
the point proton and neutron densities (see Eq. (\ref{rhoW})).
The resulting Coulomb potential $V(r)$ and weak potential 
$A(r)$ (see Eq.(\ref{pot2})) are used to construct $U_\pm(r)$.
The cross sections for elastic electron scattering are 
obtained from  the 
numerical solution of the Dirac equation for electron scattering
in the $U_\pm(r)$ potential
and includes Coulomb distortion effects \cite{Rufa1982273,Moreno2009306,
PhysRevC.57.3430,PhysRevC.72.044307}.
The cross sections for positive and negative helicity electron 
states are calculated and the resulting asymmetry parameter 
$A_{pv}$ is plotted as a function of the scattering angle.

\begin{figure}
\begin{center}
\vskip .8cm
\includegraphics[scale=0.35,angle=0.0]{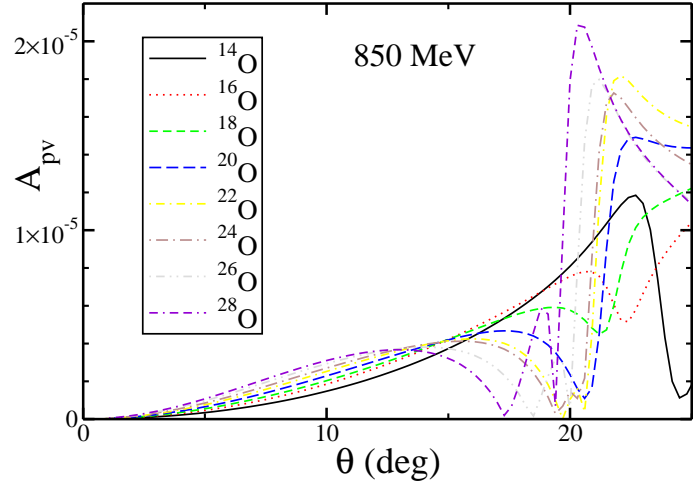}
\end{center}
\caption{\label{figO}
(Color online) Parity violating asymmetry parameter $A_{pv}$ for elastic electron 
scattering at $\varepsilon  = 850$ MeV as function of the scattering angle 
$\theta$ on $^{14-28}$O. 
}
\end{figure}

\begin{figure}
\begin{center}
\includegraphics[scale=0.35,angle=0.0]{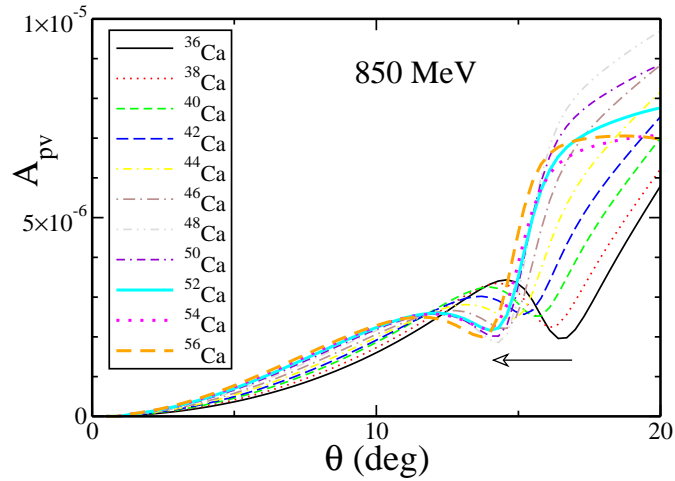}
\end{center}
\caption{\label{figCa}(Color online)   The same as in Fig.~\ref{figO}, but for $^{36-56}$Ca. 
The black arrow emphasizes the evolution of $A_{pv}$ as a function of the 
neutron number $N$.
}
\end{figure}

In Figs.~\ref{figO} and \ref{figCa}  we plot the parity-violating 
asymmetry parameters $A_{pv}$ for $^{14-28}$O and 
$^{36-56}$Ca nuclei for elastic electron scattering at $\varepsilon = 850$ MeV.
At $\varepsilon = 850$ MeV the values of $A_{pv}$ are of the order of $
10^{-5}$,  
with lower values for smaller angles and larger values for larger angles. 

\begin{figure}
\begin{center}
\includegraphics[scale=0.35,angle=0.0]{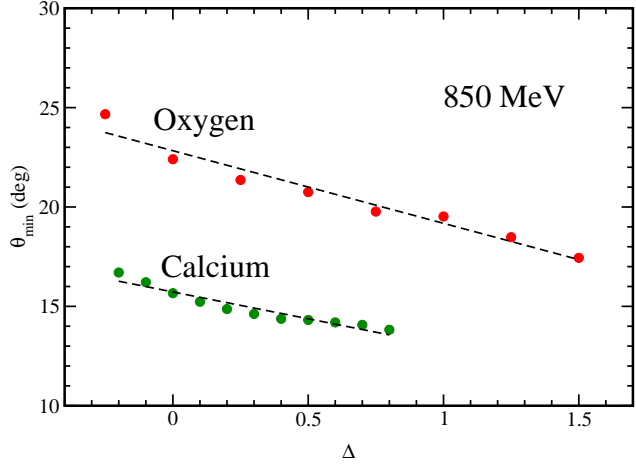}
\end{center}
\caption{\label{figocan}
(Color online) First minimum positions of the asymmetry parameter $A_{pv}$ as functions
of $\Delta =(N-Z)/Z$ for $^{14-28}$O and $^{36-56}$Ca. 
The black dashed lines represent the best linear fit. 
}
\end{figure}

\begin{figure}
\begin{center}
\includegraphics[scale=0.35,angle=0.0]{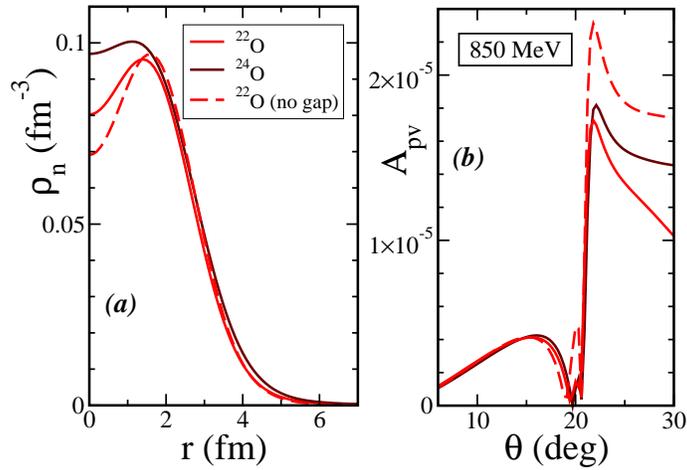}
\end{center}
\caption{\label{figbubble}(Color online)  Panel (a): neutron density distributions for 
some selected nuclei ($^{22}$O and $^{24}$O) 
that could be interpreted as candidates for \lq\lq bubble\rq\rq\ structure \cite{PhysRevC.79.034318}.
Panel (b): parity violating asymmetry parameter $A_{pv}$ for elastic electron
scattering at $\varepsilon = 850$ MeV as a function of the scattering 
angle $\theta$. 
}
\end{figure}

As suggested in Ref. \cite{Donnelly1989589} the asymmetry parameter $A_{pv}$ provides a direct measurement
of the Fourier transform of the neutron density. This relation has been tested and confirmed
in Ref. \cite{PhysRevC.61.064307} comparing asymmetries  and 
the squares of the Fourier transforms of the neutron densities.
Another way to relate $A_{pv}$ to neutron distributions of finite nuclei is by looking
at possible linear correlations between the asymmetry parameter and 
some well defined observables. We suggest to use
the first minima positions $\theta_{min}$ and the neutron excess $\Delta = \dfrac{N-Z}{Z}$, i.e.,
how the minima of $A_{pv}$ evolve from neutron-poor to 
neutron-rich nuclei (see Eqs. (\ref{AP}) and (\ref{AP2})).
In Fig.~\ref{figocan}, $\theta_{min}$ is plotted as a function of $\Delta$ for oxygen  and calcium
 isotopes. The dashed lines suggest that for both isotope chains the evolution of
$A_{pv}$ as function of $\Delta$ is well approximated by a linear fit with a very similar slope.
To test the robustness of this correlation it is interesting to study 
if $A_{pv}$ is affected by density distribution oscillations
at small radii that could appear in some selected cases. In Ref. \cite{PhysRevC.79.034318},
$^{22}$O and $^{24}$O isotopes have been studied as possible candidates for
\lq\lq bubble\rq\rq\ nuclei, i.e., nuclear systems 
with a strong depleted central density.
In Fig. \ref{figbubble} we plot the asymmetry parameters $A_{pv}$ for these 
nuclei.  
Neutron density profiles show large differences at small distances. 
No appreciable effects are obtained in the corresponding 
asymmetries up to $\theta \simeq 20^{\mathrm{o}}$, then for larger scattering 
angles the asymmetries are sensitive to the differences in the density 
distributions and are significantly different.
Therefore, $A_{pv}$ is still a reliable observable to study neutron radii 
even if we include pairing correlations, but 
we must limit to angles smaller than the first minimum position.

\begin{figure}
\begin{center}
\includegraphics[scale=0.35,angle=0.0]{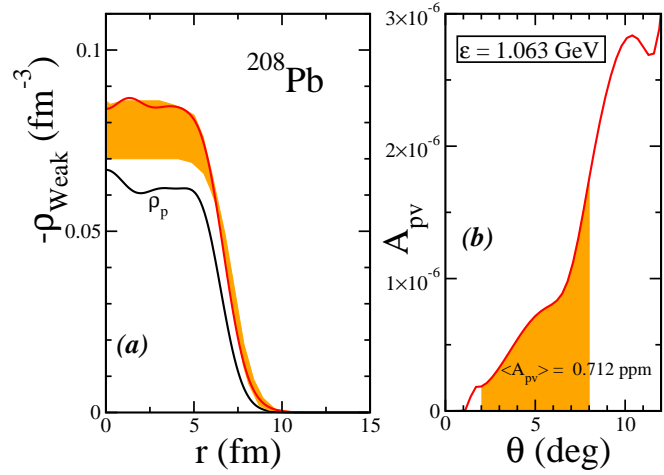}
\end{center}
\caption{\label{figApv208}
(Color online)  Panel (a): Theoretical weak charge density (red line) in 
comparison with the experimental 
error band as determined in Ref. \cite{PhysRevC.85.032501} for $^{208}$Pb with the kinematics
of the PREX experiment.
The corresponding proton density profile is plotted with a black line. 
Panel (b): Asymmetry parameter for $^{208}$Pb as a function of the angle $\theta$ (red line). 
The shaded area
represents the interval covered by the acceptance function $\epsilon(\theta)$ 
(see Ref. \cite{Abrahamyan:2012gp}
for more details). 
The asymmetry parameter averaged over the acceptance 
$\langle A_{pv} \rangle$ is in quite good agreement with the empirical 
value 0.656 $\pm$ 0.060(stat) $\pm$ 0.014(syst) ppm. 
}
\end{figure}

\begin{figure}
\begin{center}
\end{center}
\includegraphics[scale=0.35,angle=0.0]{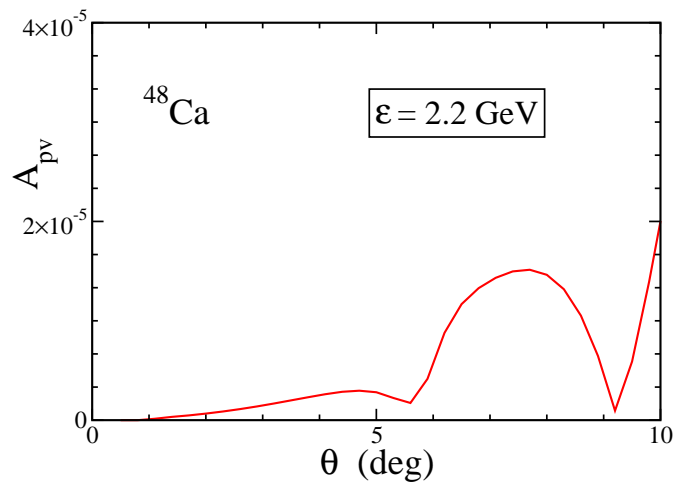}
\caption{\label{figApv48} (Color online) Parity violating asymmetry parameters $A_{pv}$ for 
elastic electron scattering at $\varepsilon = 2.2$ GeV as a function of the 
scattering angle $\theta$ for $^{48}$Ca, see Ref. \cite{crex}.
}
\end{figure}

In addition to predictions about oxygen and calcium isotopic chains we also 
provide calculations for recent measurements and future experiments.
In Fig. \ref{figApv208} we show our theoretical predictions for the empirical 
values extracted from the first 
run of the PREX experiment on $^{208}$Pb at $\varepsilon = 1.06$ GeV.
In Ref. \cite{PhysRevC.85.032501} the weak charge 
density $(-\rho_W)$ has been deduced from the weak charge form 
factor. The error band (shaded area) represents the incoherent 
sum of experimental and model errors. Our prediction, plotted by 
the red line in the left panel, is in rather good agreement with 
empirical data. In fact, if we evaluate the corresponding asymmetry 
parameter $A_{pv}$ averaged over the acceptance 
function $\epsilon(\theta)$ \cite{acceptance}
\begin{equation}
\langle A_{pv} \rangle = \dfrac{\int {\rm d}\theta~ \sin \theta A_{pv}(\theta) 
\dfrac{d\sigma}{d\Omega} \  \epsilon(\theta)}
{\int {\rm d}\theta~ \sin \theta \ \dfrac{d\sigma}{d\Omega} \ \epsilon(\theta)}
\end{equation}
we find 0.712 ppm, in very good agreement with the empirical estimate 0.656 $\pm$ 0.060(stat) 
$\pm$ 0.014(syst) ppm.
In Fig. \ref{figApv48} we calculate the asymmetry parameter for $^{48}$Ca with 2.2 GeV 
electrons as planned
for the CREX experiment. For energies well above the 1 GeV 
region, of course, the elastic scattering approximation is
not completely under control and corrections due to possible 
inelasticities should be taken into account. We plan to 
extend our calculation in a forthcoming paper.

\begin{figure}[tb]
\begin{center}
\end{center}
\includegraphics[scale=0.45]{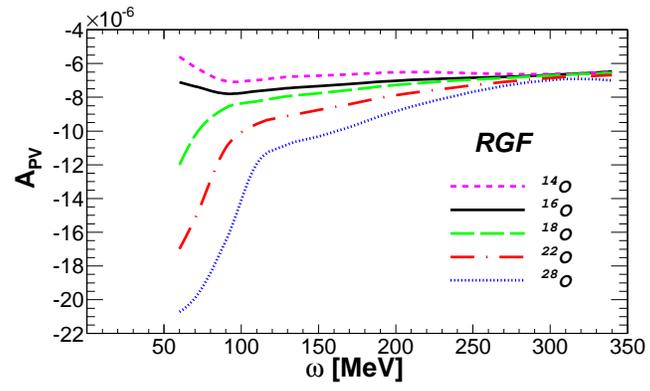} 
	\caption{(Color online) Parity-violating asymmetry for the quasielastic 
	$(e,e^{\prime})$  reaction on $^{}$O isotopes with the relativistic 
	Green\rq{}s function
	for  $\varepsilon = 1080$ MeV and  
	$\theta = 32^{\mathrm{o}}$.
	\label{foiso-apv-gf}}
	\end{figure}

As an example of the parity-violating asymmetry for QE scattering, in Fig.~\ref{foiso-apv-gf} we show 
$A_{pv}$ for the $^{14,16,18,22,28}$O isotopes evaluated with the RGF in the same kinematics as in  
Fig.~\ref{foiso-gf}. Note that the results are 
rescaled by the factor $10^5$. The RPWIA results are always similar to the 
RGF ones and are not presented here.   
The asymmetry is almost constant to a few $\times 10^{-6}$ 
for $^{14}$O and $^{16}$O, whereas for $^{22}$O and $^{28}$O it goes up to 
$\approx -2\times 10^{-5}$ 
 in the low energy transferred region. There are visible relative differences 
between the asymmetries for the oxygen isotopes that we have considered, but we 
are aware that a measurement 
of $A_{pv}$ in QE electron scattering off finite nuclei is extremely challenging 
with the presently available  facilities.


\section{Summary and conclusions}
\label{conc}

We have presented and discussed numerical predictions for the 
cross section and the parity-violating asymmetry in elastic and quasi-elastic 
electron scattering on oxygen and calcium isotope chains with the aim to
investigate their evolution with increasing neutron number. 

The understanding of the properties of exotic nuclei is one of the major topic of interest in 
modern nuclear physics.   
Large efforts in this directions have been done over last years and are 
planned for the future. The use of electrons as probe provides a powerful 
tool to achieve this goal.
The  RIB facilities in different laboratories have opened the possibility to give 
insight into  nuclear structures which are not available in nature, as they
are not stable, but which are important in astrophysics and had a relevant 
role in the nucleosynthesis.

Electron scattering is well fitted for studying nuclear properties, as its 
interaction is well known and relatively weak with respect to the hadron force 
and can therefore more adequately explore the details of inner nuclear structures. 
As a consequence of this weakness, the cross sections become very small and more 
difficult experiments have to be performed. Electron scattering experiments off
exotic nuclei have been proposed in the ELISe experiment at FAIR and in the
SCRIT project at  RIKEN. We hope that the existing 
proposals will be considered and approved in next years. Our 
theoretical predictions will be useful for clarifying the different aspects of 
the measurements, giving information on the order of magnitude of the 
measurable quantities and therefore making possible a more precise evaluation 
of the experimental difficulties. Moreover, a theoretical investigation can 
be helpful to envisage the most interesting quantities to be measured 
in order to explore the properties of exotic nuclear structures.

In this work, both elastic and inclusive quasi-elastic electron scattering have
been considered. The elastic scattering can give information on the 
global properties of nuclei and, in particular, on the different behavior of 
proton and neutron density distributions. The inclusive quasi-elastic scattering 
is  affected by the dynamical properties, being the integral of the spectral 
density function over all the available final states, and, due to the reaction 
mechanism, preferably exploits the single particle aspects of the nucleus. 
In addition, when combined with the exclusive $(e,e'p)$ scattering, it is able 
to explore the evolution of the single particle model with increasing asymmetry
between the number of neutrons and protons. 
Many interesting phenomena are predicted in this 
situation: in particular, the modification of the shell model magic numbers. 
A definite response can be obtained from the comparison with experimental 
data, which will discriminate between the different theoretical models, mainly 
referring to relativistic mean field approaches.

As case studies for the present investigation we have selected oxygen and
calcium isotope chains.  The calculations have been carried out within the 
framework 
of the relativistic mean field model. The nuclear wave functions are obtained 
considering a system of nucleons coupled to the exchange mesons and the 
electromagnetic field through 
an effective Lagrangian. The calculated cross sections include both the hadronic 
and Coulomb final states interactions. The inclusive quasi-elastic scattering 
is calculated with the relativistic Green's function model, which conserves the 
global particle flux in all the final state channels, as it is required in an 
inclusive reaction.

First, the models have been compared with experimental data 
already available on stable isotopes in order to check their reliability. Then, 
the same models have been used to calculate elastic and inclusive quasi-elastic
cross sections on exotic isotopes chains. 
The possible disagreement of the experimental findings from the  theoretical 
predictions will be a clear indication of the insurgence of new phenomena 
related to the proton to neutron asymmetry. 

Our results show an evolution of the calculated quantities without 
discontinuities. 
The increase of the neutron number essentially produces an increase of the 
nuclear and proton densities and a flattening of the charge density.

The parity-violating asymmetry parameter has been calculated in order to 
investigate the neutron skin, as the weak current is essentially obtained 
from the interaction with neutrons. Numerical predictions have been presented
and discussed for oxygen and calcium isotopes. Moreover, we have compared
our calculations with the results of the first measurement of the asymmetry 
parameter achieved by the PREX Collaboration on \pbd\ and have obtained a good 
agreement with the empirical value. 
A prediction for the future experiment CREX on \caq\ has also been given.
\begin{acknowledgments}

This work was partially supported by the Italian MIUR 
through the PRIN 2009 research project.

\end{acknowledgments}

%

\newpage


\end{document}